# Fermi National Accelerator Laboratory
PARTI 2011
Report
## Measurement of magnetic permeability of steel laminations of Booster gradient magnets

Yury Tokpanov
supervisors: Valeri Lebedev, William Pellico

**Abstract**
*New experiments at Fermilab require higher beam intensities at Booster, resulting in higher influence of magnet steel on the beam. In order to carefully take these effects into account it is necessary to know geometric and material properties of magnets. This work is dedicated to the measurement of magnetic permeability of steel laminations of Booster gradient magnets as a complex function of frequency in the frequency range from several megahertz up to 1 GHz. The magnetic permeability is estimated by analyzing S-parameters of microstrip lines made up from laminated steel.*

## 1. Background

Booster is mainly comprised of so-called gradient magnets. These magnets have the combined function of bending and focusing the beam. Basically, Booster gradient magnets are dipole magnets with the magnetic field gradient, which is achieved by using special geometry (see the fig. 1).

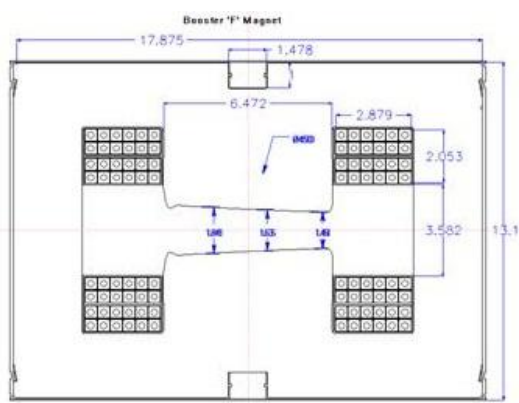

Fig.1 Geometry of Booster gradient magnets

When the beam propagates through the chamber, it induces currents in the walls of the chamber. The field of these induced currents influences the motion of the beam, creating instabilities. Given known the geometry and material properties of the chamber, these effects can be taken into account.

Usually, the beam is placed inside a pipe of a known geometry and material properties. Since no beampipe is used in Booster gradient magnets, one should know the properties of the Booster gradient magnets materials.

While geometry is known for some degree, the magnetic permeability of the laminated steel, which is used in these magnets, is unknown in desired frequency range.

## 2. Objective
The objective of this project is to measure the magnetic permeability of laminated steel used in Booster gradient magnets as a complex function of frequency in the frequency range from several megahertz up to 1 GHz.

## 3. Approach

### 3.1 Idea of the measurement

The propagation of electromagnetic waves depends upon properties of the materials, in which they propagate.

The main idea is to make a microstrip (or strip) line (see fig.2) and measure how electromagnetic waves reflect and transmit through such system.

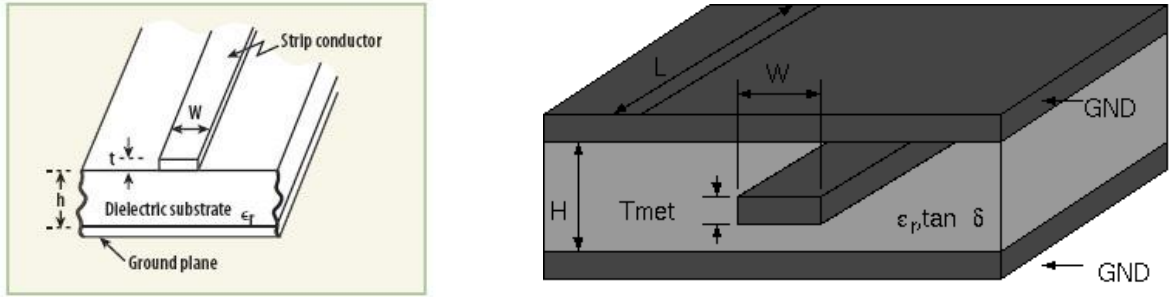

Fig.2 Microstrip (on the left) and strip (on the right) lines.

The magnetic permeability determines the resistive losses of a transmission line, so, by measuring the effects, caused by them, it is possible to estimate the magnetic permeability as a complex function of frequency.

### 3.2 1D transmission line model

One-dimensional transmission line can be represented by an effective scheme, which comprises of series of infinitesimally small elements consisting of resistance, inductance and capacitance per unit length, as shown on fig. 3.

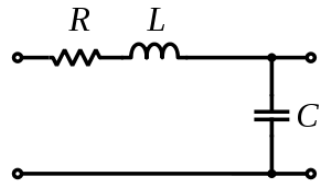

Fig.3 Effective scheme of infinitesimal small element of the transmission line

By applying usual Ohm's laws to the above scheme the following expressions for voltage and current distributions (so-called telegrapher's equations) can be obtained:

$$\begin{cases} \dfrac{\partial U}{\partial x} = -IR - L\dfrac{\partial I}{\partial t} \\ \dfrac{\partial I}{\partial x} = -C\dfrac{\partial U}{\partial t} \end{cases} \quad (1)$$

These equations have harmonic solutions (two terms with opposite sign of $k$ correspond to two different directions of wave in the transmission line):

$$U = Ae^{j\omega t - jkx} + Be^{j\omega t + jkx}$$
$$I = \dfrac{Ajk}{R + j\omega L}e^{j\omega t - jkx} - \dfrac{Bjk}{R + j\omega L}e^{j\omega t + jkx}, \quad (2)$$

where $k^2 = -j\omega C(R + j\omega L)$ is a wave number, $A$ and $B$ are coefficients determined by boundary conditions, $j$ is an imaginary unit.

From (2) one can easily obtain characteristic impedance of the transmission line:

$$\rho = \frac{U}{I} = \sqrt{\frac{R + j\omega L}{j\omega C}}, \qquad (3)$$

$k$ and $\rho$ are important parameters, which almost fully determine the propagation of current and voltage waves in the transmission line.

### 3.3 Parameters of transmission line

Assuming the width of a strip conductor in a microstrip line to be much larger than a dielectric thickness, one can obtain the following simple expressions for capacitance, inductance and resistance per unit length:

$$C = \varepsilon\varepsilon_0 \frac{W}{H}, \quad L = \mu_0 \frac{H}{W}, \quad R = \frac{1+j}{W}\left[\left(\sqrt{\frac{\omega\mu\mu_0}{2\sigma}}\right)_{strip\ conductor} + \left(\sqrt{\frac{\omega\mu\mu_0}{2\sigma}}\right)_{ground\ conductor}\right], \qquad (4)$$

where $W$ is a width of a strip conductor, $H$ – a thickness of a dielectric, $\varepsilon$ – dielectric constant, $\mu$ – magnetic permeability, $\omega$ – angular frequency, $\sigma$ – conductivity, $\mu_0 = 4\pi \cdot 10^{-7}$ H/m, $\varepsilon_0 = 8.854187817620... \cdot 10^{-12}$ F/m.

Capacitance and inductance are fully determined by the geometry of a microstrip (strip) line and dielectric constant.

In literature more complicated formulae exist, which take into account edge effects [1-3]. Such formulae can be obtained numerically and analytically (for example, by using conformal mapping into a region, where the width of a strip conductor is much larger than a dielectric thickness, and then by using expressions (4)). However, we don't need them since characteristic impedance and wave number in case of negligibly small resistive losses can be obtained experimentally and then used for analyzing steel microstip lines (as will be shown below).

Both dielectric constant and magnetic permeability have complex values:

$$\varepsilon = \varepsilon' - i\varepsilon'', \quad \mu = \mu' - i\mu'' \qquad (5)$$

Imaginary parts correspond to losses. The following parameter is called dielectric loss tangent:

$$\tan\delta = \frac{\varepsilon''}{\varepsilon'} \qquad (6)$$

Usually, this parameter is pretty small in the sense that allows one to drop tangent in (6). Then dielectric constant can be written as:

$$\varepsilon = \varepsilon'(1 - j\delta) \qquad (7)$$

### 3.3.1 Characteristic impedance and wave number in the case of negligibly small resistive losses.

For copper, in which resistive losses are negligible because of high conductivity (around 60 S/m), one can drop out $R$ in formulae for characteristic impedance and wave number:

$$\rho_c \approx \sqrt{\frac{L}{C}} = \sqrt{\frac{L}{C'(1-j\delta)}} \approx \rho_0 (1 + j\delta/2), \qquad (8)$$

$$k_c l \approx \omega l \sqrt{CL} = \omega l \sqrt{C'(1-j\delta)L} \approx \omega \tau_c (1 - j\delta/2),$$

where $\rho_0 = \sqrt{\frac{L}{C'}}$ and $C'$ are characteristic impedance and capacitance per unit length respectively in the case of zero dielectric losses, $\tau_c = l\sqrt{LC'}$ is a time delay in the transmission line, $l$ is a length of a transmission line, subscript $c$ stands for copper.

It is worth noting, that $\rho_0$ and $\tau_c$ in this case are fully determined only by geometry of a microstrip and dielectric constant of a dielectric.

*3.3.2 Basic formulae for steel strips analysis.*

In the case of steel, where resistive losses are high and cannot be neglected, one can easily obtain the following expressions (if the geometry and the dielectric are the same with copper microstrip line):

$$\rho_s = \sqrt{\frac{R + j\omega L}{j\omega C}} = \sqrt{\frac{L}{C}\left(1 + \frac{R}{j\omega L}\right)} = \rho_c \sqrt{1 + \frac{R}{j\omega L}}, \qquad (9)$$

$$\tau_s = \tau_c \sqrt{1 + \frac{R}{j\omega L}},$$

where $R$ is the steel resistance per unit length, subscript $s$ stands for steel, $c$ – for copper.

Inductance per unit length in (9) can be calculated from $\rho_0$ and $\tau_c$:

$$L = \frac{\rho_0 \cdot \tau_c}{l} \qquad (10)$$

One of the important questions is what formula for $R$ is suitable for our analysis. Rough estimations of magnetic permeability can be done by using formula (4) (one can drop the ground conductor term, since in all our experiments it was copper, which resistance is very small in comparison with steel). In [4] there is a following approximate formula for resistance per unit length of a strip conductor in a microstrip line (obtained by conformal mapping technique):

$$R = (1+j)\frac{LR}{W}\sqrt{\frac{\omega\mu\mu_0}{2\sigma}}\left(\frac{1}{\pi} + \frac{1}{\pi^2}\ln\frac{4\pi W}{T}\right),$$

$$\text{where } LR = \begin{cases} 1, \text{ if } \frac{W}{H} \leq 0.5, \\ 0.94 + 0.132\frac{W}{H} - 0.0062\left(\frac{W}{H}\right)^2, \text{ if } \frac{W}{H} > 0.5, \end{cases} \qquad (11)$$

where $W$ is a width of a strip conductor, $T$ – thickness of a strip conductor, $H$ – thickness of a dielectric. This formula works pretty well for $W/H$ up to the order of several decades (which is the case in our experiments). The order of resistance, calculated using it, is the same with the resistance, calculated using (4).

For magnetic permeability formula, based on phenomenological Landau-Lifshitz ferromagnetic resonance (LL FMR) model, gives good qualitative agreement with real data [5], that's why it is useful to make a fitting of the parameters of this formula:

$$\mu = 1 + \frac{\mu_s}{1 + j\frac{f}{f_a} - \left(\frac{f}{f_r}\right)^2}, \qquad (12)$$

where $f$ is a frequency, $\mu_s$, $f_a$ and $f_r$ are parameters, which are based on other intrinsic parameters of LL FMR model, we used these parameters for fitting the model to experimental data.

### 3.4 S-parameters

Parameters, mentioned above, can be estimated by analyzing so-called S-parameters of a transmission line. S-parameters characterize scattering of electromagnetic waves on the transmission line. If $a_1$, $a_2$ are amplitudes of incident signals and $b_1$, $b_2$ are amplitudes of scattered signals (see fig. 4), then S-parameters are defined as follows:

$$\begin{pmatrix} b_1 \\ b_2 \end{pmatrix} = \begin{pmatrix} S_{11} & S_{12} \\ S_{21} & S_{22} \end{pmatrix} \begin{pmatrix} a_1 \\ a_2 \end{pmatrix}, \qquad (13)$$

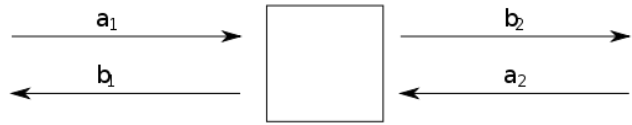

Fig.4 Incident, reflected and transmitted waves (this accompanies the definition of S-parameters).

Equations for S-parameters can be derived by satisfying boundary conditions for voltage and current. Schematics, which helps writing down necessary expressions, is represented in the fig. 5. The following system of equations can be obtained:

$$\begin{aligned} U_0 + U_1 &= U_2 + U_3 \\ \frac{U_0 - U_1}{Z_0} &= \frac{U_2 - U_3}{\rho} \\ U_2 e^{-jkl} + U_3 e^{jkl} &= U_4 \\ \frac{U_2 e^{-jkl} - U_3 e^{jkl}}{\rho} &= \frac{U_4}{Z_0} \end{aligned} \qquad (14)$$

where $\rho$ is a characteristic impedance of a transmission line, $Z_0 = 50\Omega$ is an ideal characteristic impedance of measuring system, $l$ – length of the line, $k$ – wave number, meanings of voltages amplitudes are presented on the fig. 5.

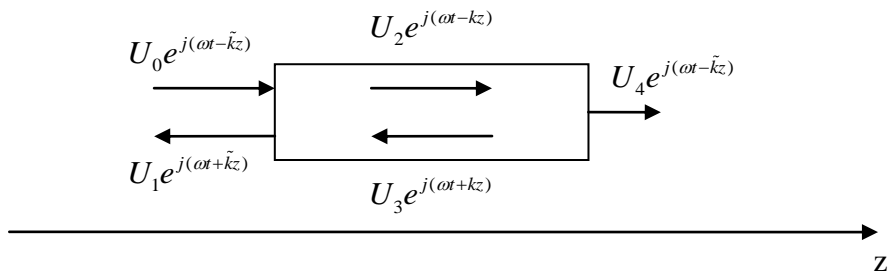

Fig.5 Voltage waves in the system

By solving this system one can obtain the following expressions for S-parameters:

$$S_{11} = \frac{U_1}{U_0} = \frac{j(\kappa^2 - 1)\tan kl}{2\kappa + j(\kappa^2 + 1)\tan kl}$$
$$S_{21} = \frac{U_4}{U_0} = \frac{2\kappa}{2\kappa \cos kl + j(\kappa^2 + 1)\sin kl}$$

(15)

where $\kappa = \frac{\rho}{R}$.

By fitting formulae (8)-(15) to experimental data it is possible to estimate all needed parameters.

## 4. Results

All measurements were done by using Agilent E5061B Network Analyzer, which was calibrated with 85033D calibration kit. Obtained data were analyzed with the help of MATLAB.

### 4.1 Copper measurements

Before making measurements on steel it is necessary to carry out analysis on microstrip lines made up of copper.

During analysis of copper microstrip lines, additional phase shifts to reflected and transmitted waves were found (see fig. 6).

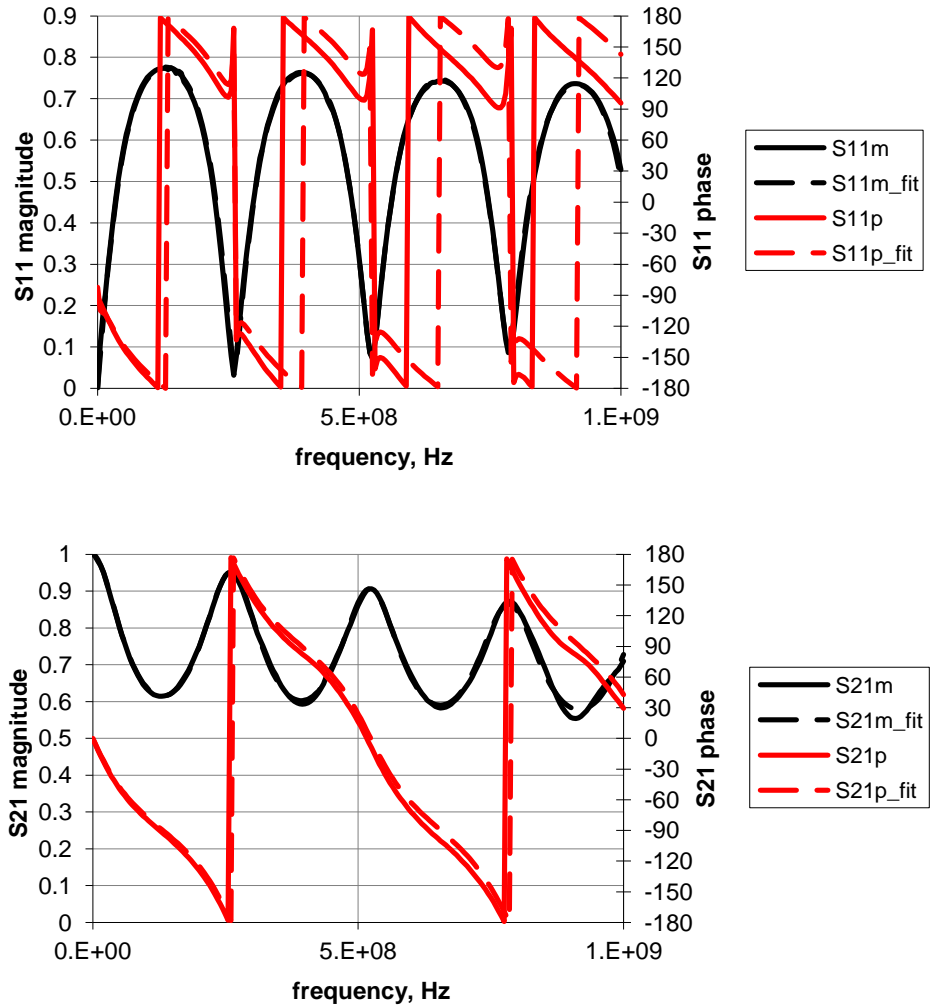

Fig.6 Manifestation of additional phase shifts on thin copper microstrip line with $W \approx 12$mm, $H \approx 1.4$mm.
Determined parameters are: $\rho_c = 17.4\,\Omega$, $\tau_c = 1.91 \cdot 10^{-9}$ s, $\delta = 0.02$

From this picture it can be seen that fitting to a magnitude can be done in a very good manner. In addition, the phase shape of the fitting model is pretty much the same as the experimental one.

This additional phase shift can be attributed to several reasons, including:
1. Not proper theoretical model:
    a. radiation from pins of SMA connectors can cause an additional phase shift;
    b. different widths of SMA pins and microstrip cause a transition region between pin current and microstrip current. Such region can effectively work as an increase in length of the connector. This effect can be taken into account by solving 2D analogue of telegrapher's equations with proper boundary conditions.
2. Network analyzer calibration errors.

Radiation as a reason can be tested by measuring S-parameters a strip line (fig. 7). If this is the case, then, connector pin having been shielded by ground conductors from both sides of the strip, the additional phase shift should be different from the case of a microstrip line. But experiment didn't reveal such a difference (see fig. 7), which implies an exclusion of this suggestion. In addition, this experiment also shows the influence of edge effects: instead of dropping by 2 times, wave impedance decreased from 17.4Ω to 10.1Ω.

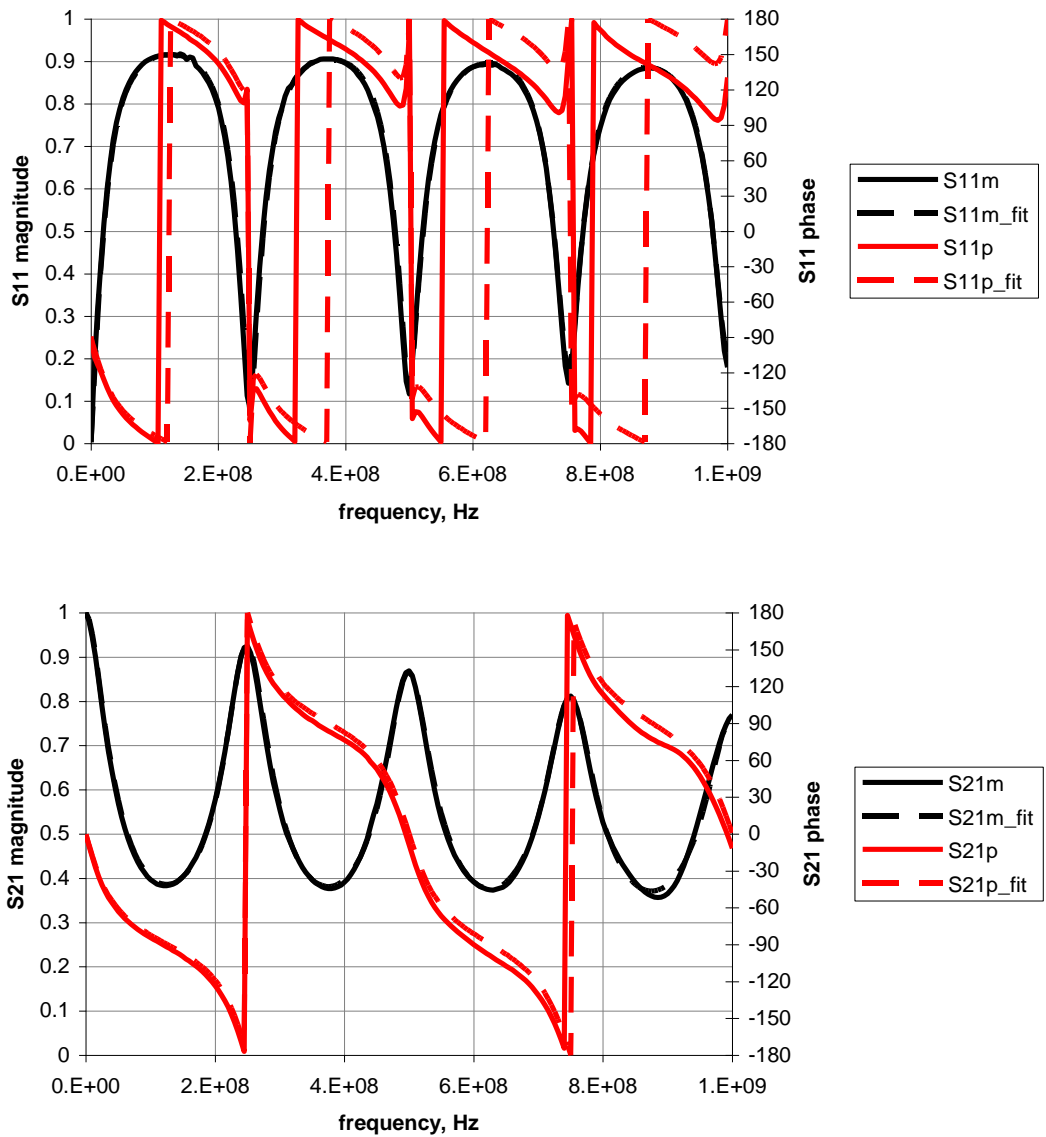

Fig.7 Manifestation of additional phase shifts on thin copper strip line with $W≈12$mm, $H≈2·1.4$mm. Determined parameters are: $\rho_c = 10.1\Omega$, $\tau_c = 2·10^{-9}$ s, $\delta = 0.02$

Another reason, transition from coaxial line to microstrip line, was verified by taking S-parameters of copper microstrip lines with different widths. No dependence of the additional phase shift on width was found.

The very simple explanation of this phase shift suggests that it is caused by the additional length in measuring system, introduced by the length of the connectors. This hypothesis is supported by the linear increase of the phase shift with the frequency. However, phase shifts for S11 and S21 must be the same under such conditions, but they are different (see fig.6 or 7). Fortunately, this difference can be explained by the features of 2-port calibration of network analyzer (see [6]).

2-port calibration procedure consists of two steps: 1-port calibration using load, open and short standards, and thru calibration. Used calibration kit didn't contain a standard thru connector, which led to phase error in thru measurements. This can be corrected by measuring the phase shift of the thru connector, used in the calibration procedure. Thus, if the S11 phase shift equals S21 phase shift plus thru connector phase shift, then the additional phase shift can be explained by errors in network analyzer calibration. And this found to be the case: at 1 GHz the S11 phase shift was around 45 degrees, S21 phase shift – around 15 degrees, thru connector phase shift – around 30.

So, copper measurements give us a way for determining $\rho_c$, $\tau_c$, dielectric losses (see section 3.3.1 for definitions) and network analyzer errors. Fig. 11 shows an example of copper data with phase correction. This example demonstrates that even simple 1-dimensional model of a transition line can describe the behavior of the real system in the desired frequency range.

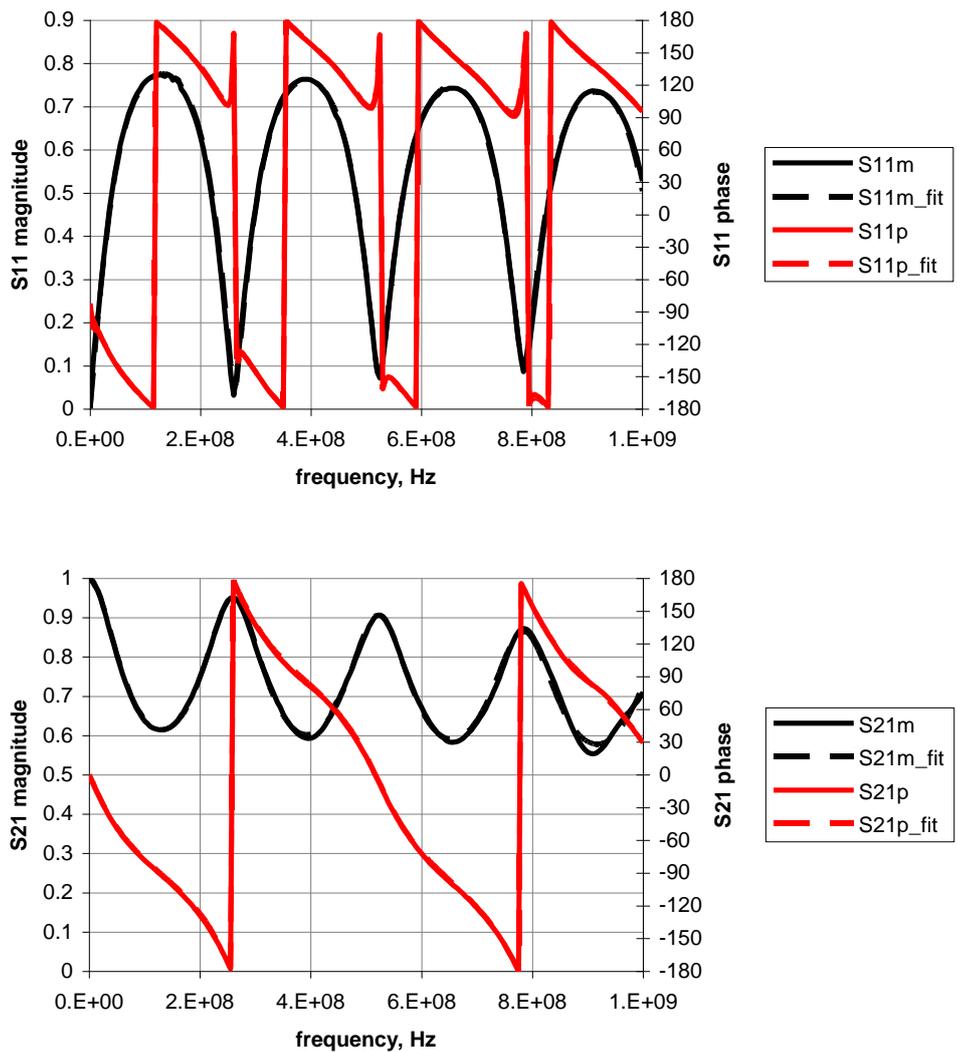

Fig.8 S-paramaters of copper microstrip line with corrected phase.

## 4.2 Steel measurements

Measurements of the steel microstrip line were carried inside the magnet at different values of the magnetic field (0T, 1T and 2T) and different orientations (parallel to the strip plane, normal to the strip plane).

For all measurements it is crucial to make a very good, tight contact between a strip conductor and a layer of a dielectric. In the case of normal magnetic field orientation the steel strip was pressed with glass of the same length by the poles of the used magnet. In parallel magnetic field measurements this couldn't be done and the steel strip was attached to the dielectric by epoxy glue.

### 4.2.1 Steel measurements in normal magnetic field

First of all, it is necessary to take measurements on copper microstrip lines in as much equal conditions, as possible. For this purpose we used copper microstrip line with the same geometry as the steel microstrip line on the same dielectric, put it into the magnet in the same way as the steel microstrip line. Actually the thicknesses of strip conductors were slightly different (0.55 mm for copper, 0.65 mm for steel), but this discrepancy should have affected the measurements very little (less than several fractions of percent) (see formula in [3] with the thickness adjustments).

It was found, that S-parameters are slightly different for copper microstrip in different magnetic fields (see fig. 9). Copper being a diamagnetic material, such difference can be caused by properties change of connectors or steel used in magnet poles (a gap of about 4 cm being between magnet and strip conductor in the experiment, one cannot totally exclude such an influence).

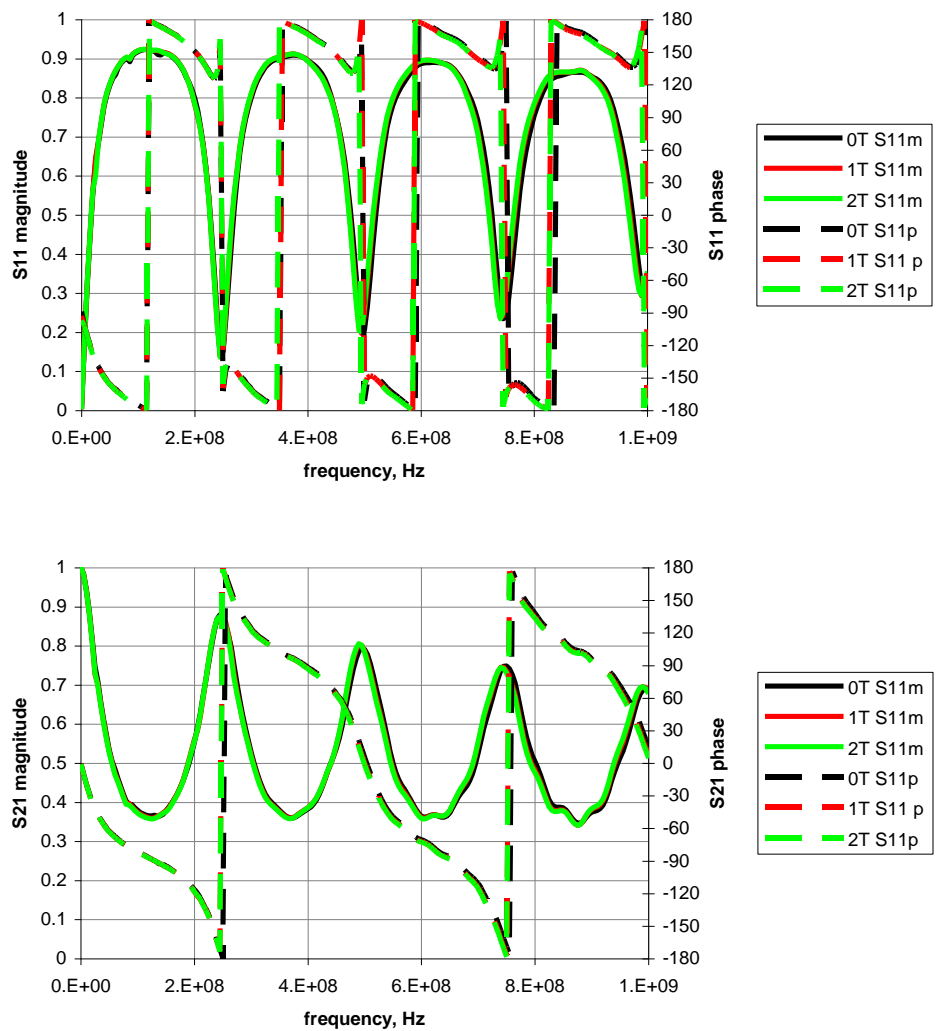

Fig.9 Influence of the magnetic field on S-parameters of the copper microstrip.

Later analysis showed that in order to increase the accuracy of a magnetic permeability calculations in different magnetic fields it was necessary to use copper data for corresponding magnetic fields. Besides, on fig.9 one can see irregularities, which increase with frequency; they are caused by longer cables, which were necessary to allow insertion of microstrip into the magnet. Such irregularities are also can be reduced by using copper data.

As anticipated, there was a noticeable difference between data of steel at different magnetic fields (see fig. 10). Transmission being higher at higher magnetic field, the resistive losses, and thus magnetic permeability, decreased.

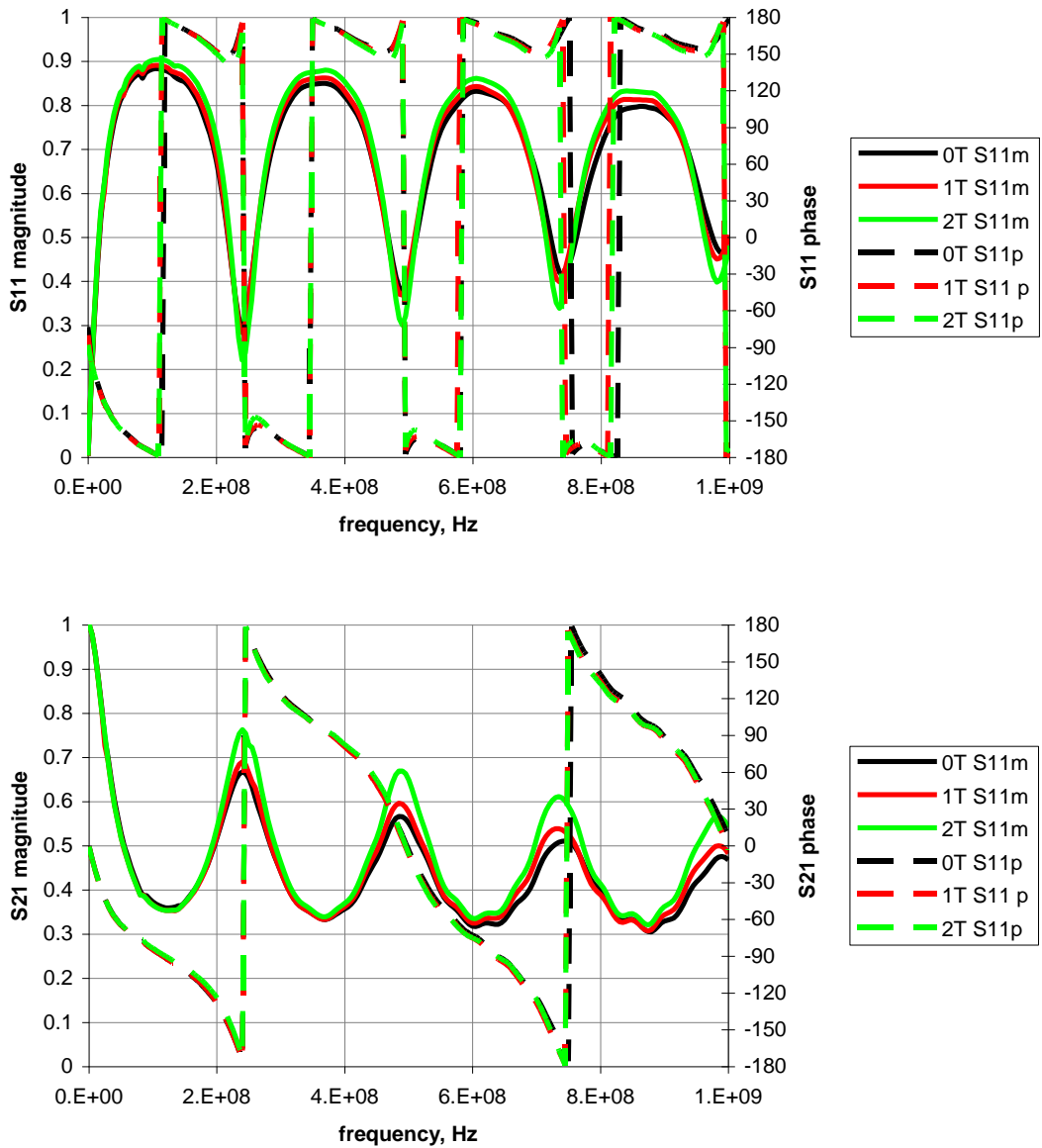

Fig.10 S-parameters of steel microstrip line in different values of magnetic field, normal to the strip plane.

As the first step, it is useful to carry out simple analysis using formula (8)-(12). In the fig. 11 one can see the fitting to copper data at 0T. Higher loss tangent can be attributed to higher losses in glass.

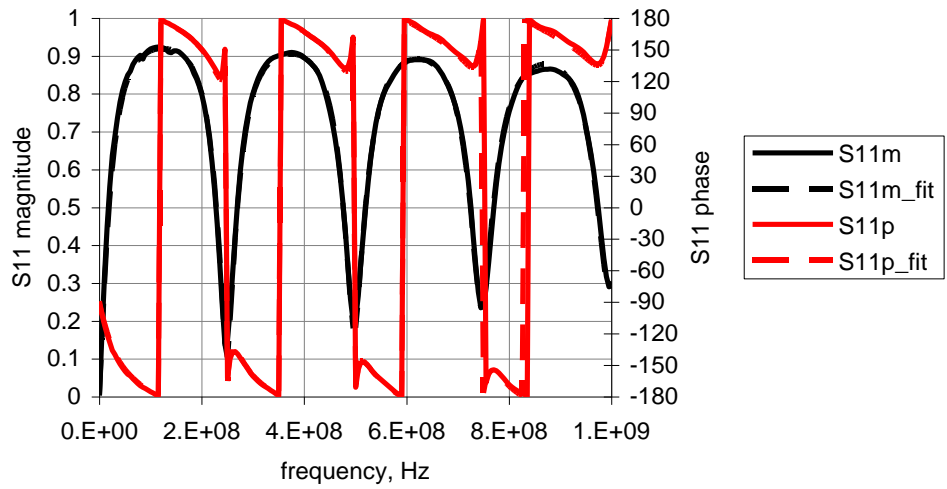

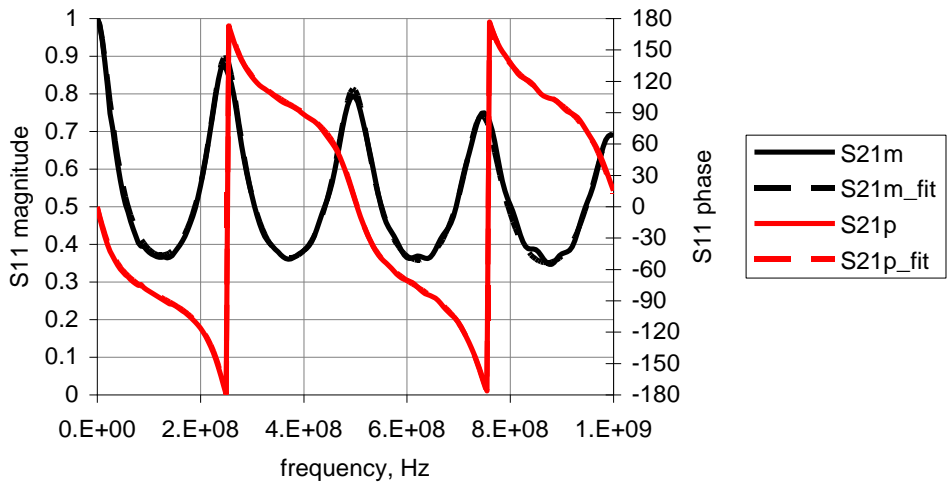

Fig.11 Reference copper microstrip line. $\rho_c = 9.7\Omega$, $\tau_c = 2.01 \cdot 10^{-9}$ s, $\delta = 0.027$

The determined parameters were used for analysis of steel microstrip. One can see the difference between S-parameters of copper and steel microstrips on the fig. 12.

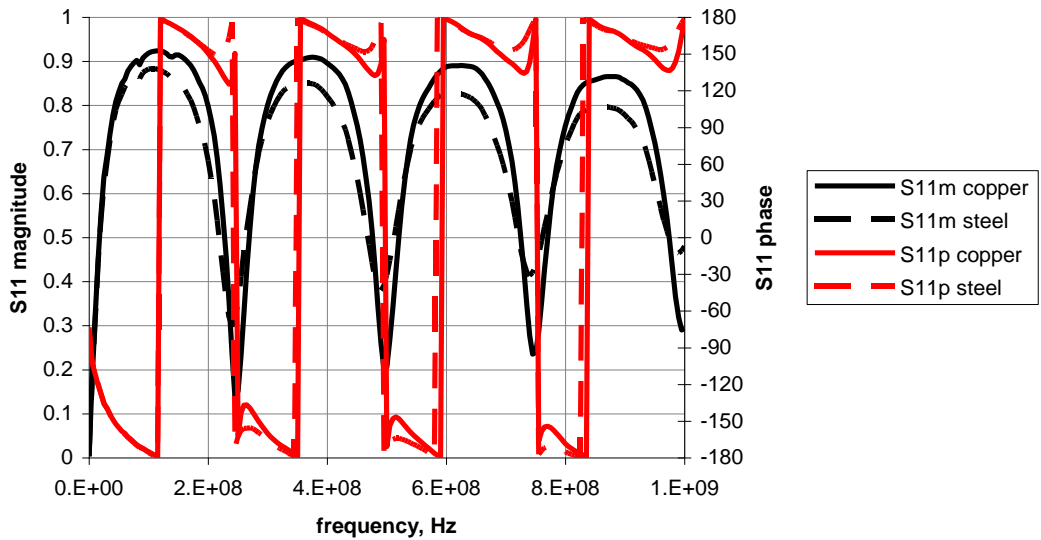

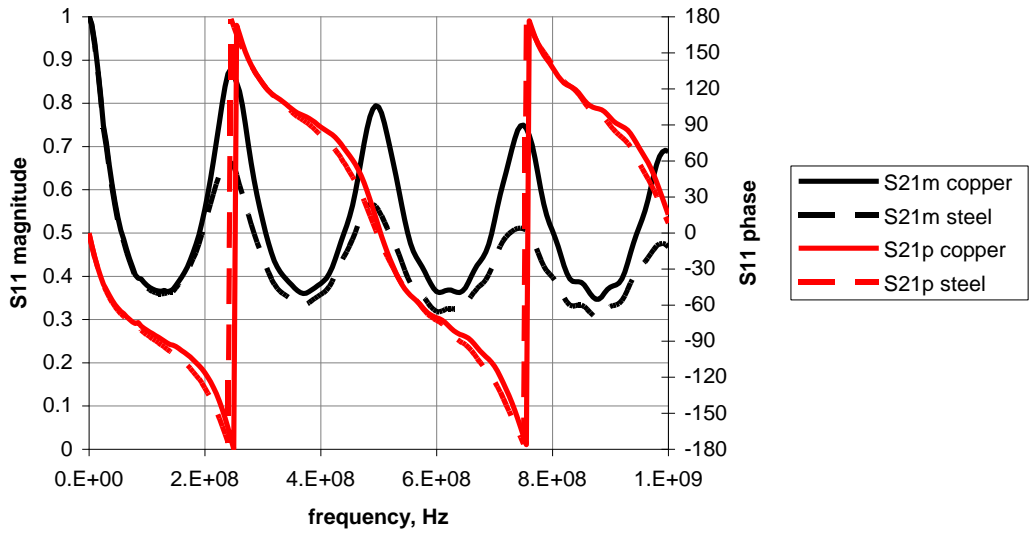

Fig.12 Comparison between S-parameters of copper and steel microstrip lines at zero magnetic field.

Fig. 13 demonstrates the fitting of LL-FMR magnetic permeability formula (12) at zero magnetic field.

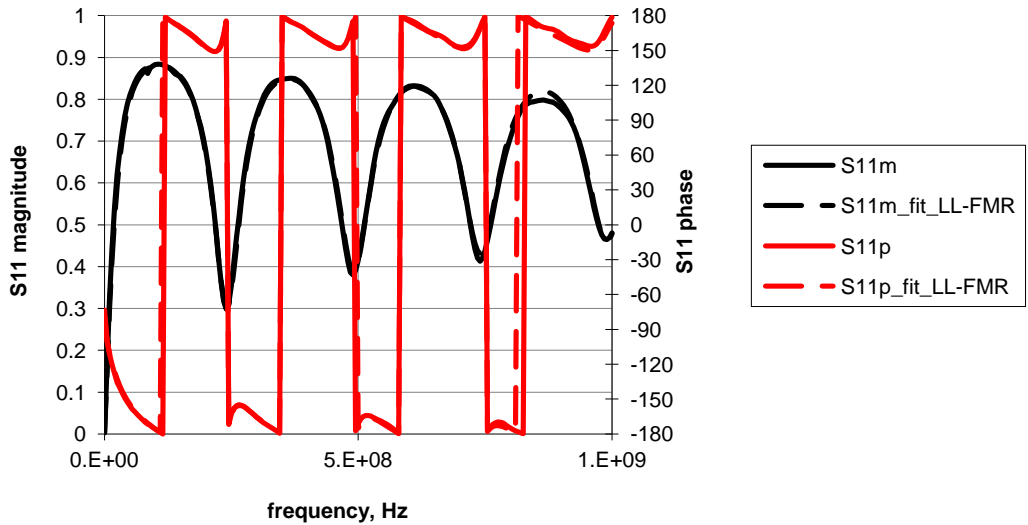

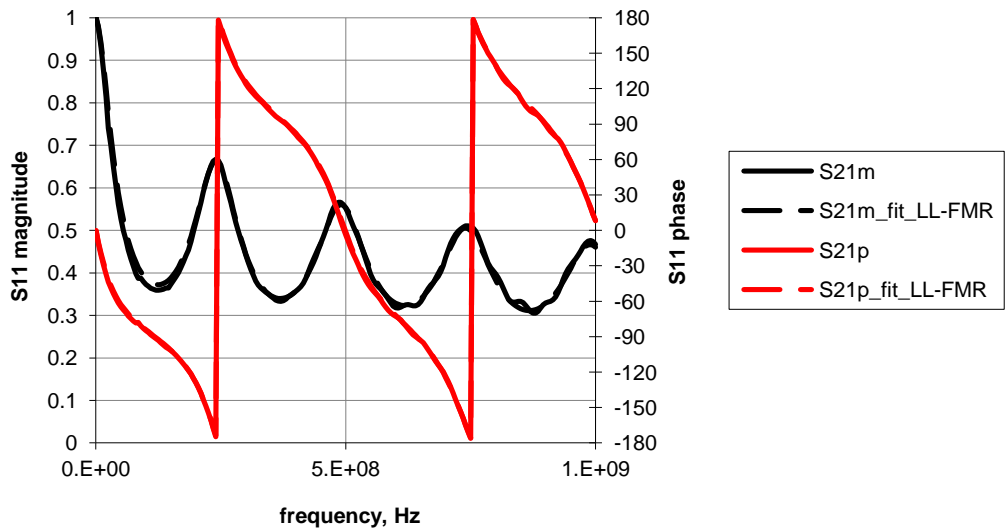

Fig.13 Fitting of LL-FMR magnetic permeability formula.

From this pictures it can be seen that fitting can be done pretty good. That implies that it is possible to determine magnetic permeability directly. For each value of frequency we have four non-linear equations for S11 and S21 (real and imaginary parts) and four parameters to determine (real and imaginary parts of wave impedance and wave number (or time delay)). Thus, by numerically solving such a system of equations it is possible to determine wave impedance and wave number as complex functions of wavelength, from which it is easy to determine magnetic permeability using formulae (8)-(11).

For convenience, it is better to rewrite formulae (15) in the following way:

$$S_{11} = \frac{U_1}{U_0} = \frac{(1-\kappa^2)(\chi^2-1)}{(1+\kappa)^2 - \chi^2(1-\kappa)^2},$$
$$S_{21} = \frac{U_4}{U_0} = \frac{4\kappa\chi}{(1+\kappa)^2 - \chi^2(1-\kappa)^2}$$
(16)

where $\kappa = \frac{\rho}{R}$, $\chi = \exp(-ikl) = \exp(-i\omega\tau)$, $k$ – wave number, $\omega$ – angular frequency, $\tau$ – time delay, $l$ – length of the transmission line.

The convenience is in the following. Trigonometric functions being periodical, the program tries determine wave number only within this period, which leads to discontinuities near the boundaries of the period; such discontinuities are hard to be handled by MATLAB solvers. However, the value $\exp(-ikl) = \exp(-i\omega\tau)$ doesn't have such problems and can be easily determined. This value being determined, it is quite easy to restore the continuous value of wave number (time delay) by adding the period after each point of discontinuity.

The fig.14 shows the calculated $\chi$ values from data of copper and steel in different magnetic field. A considerable difference between copper and steel can be seen on this picture, which is determined by high resistive losses in steel. It should be noted here, that these values were calculated on experimental data with phase corrections.

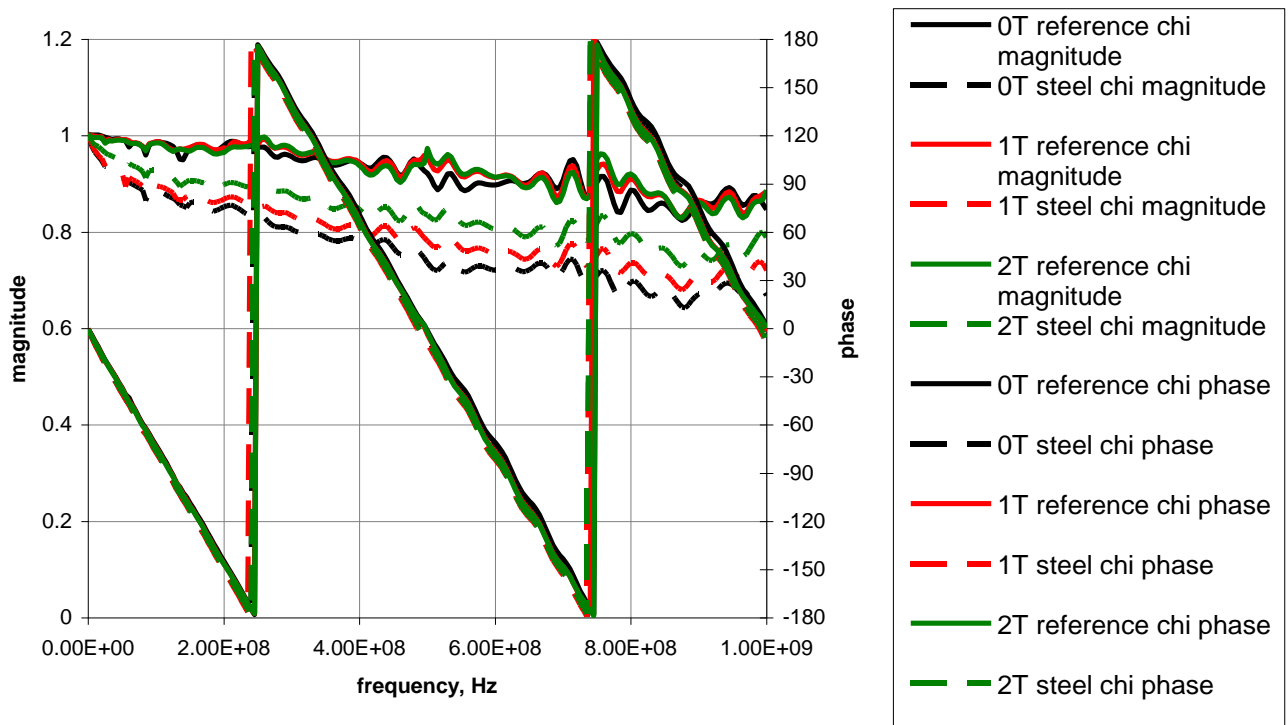

Fig.14 $\chi$ values, calculated from data of copper and steel in different magnetic field.

Irregularities on this picture are caused by measuring system errors; they have similar shape for all extracted data, which, as stated above, can be used in order to reduce them in final result. This can be achieved by simply using $\tau_c$, calculated directly from $\chi$, in steel analysis.

By using equations (9)-(10) one can calculate the steel resistance per unit length, and by using (11) the values of magnetic permeability can be obtained. The geometric parameters for steel strip: $W = 12mm$, $H = 0.75mm$, $T = 0.65mm$. In order to determine magnetic permeability, we must know conductivity σ. Since our frequencies are lower than optical by several orders, conductivity is independent on frequency and can be measured by usual dc techniques. Four-point measuring method revealed a steel conductivity of about $\sigma = 2.3 \cdot 10^6 \, S/m$.

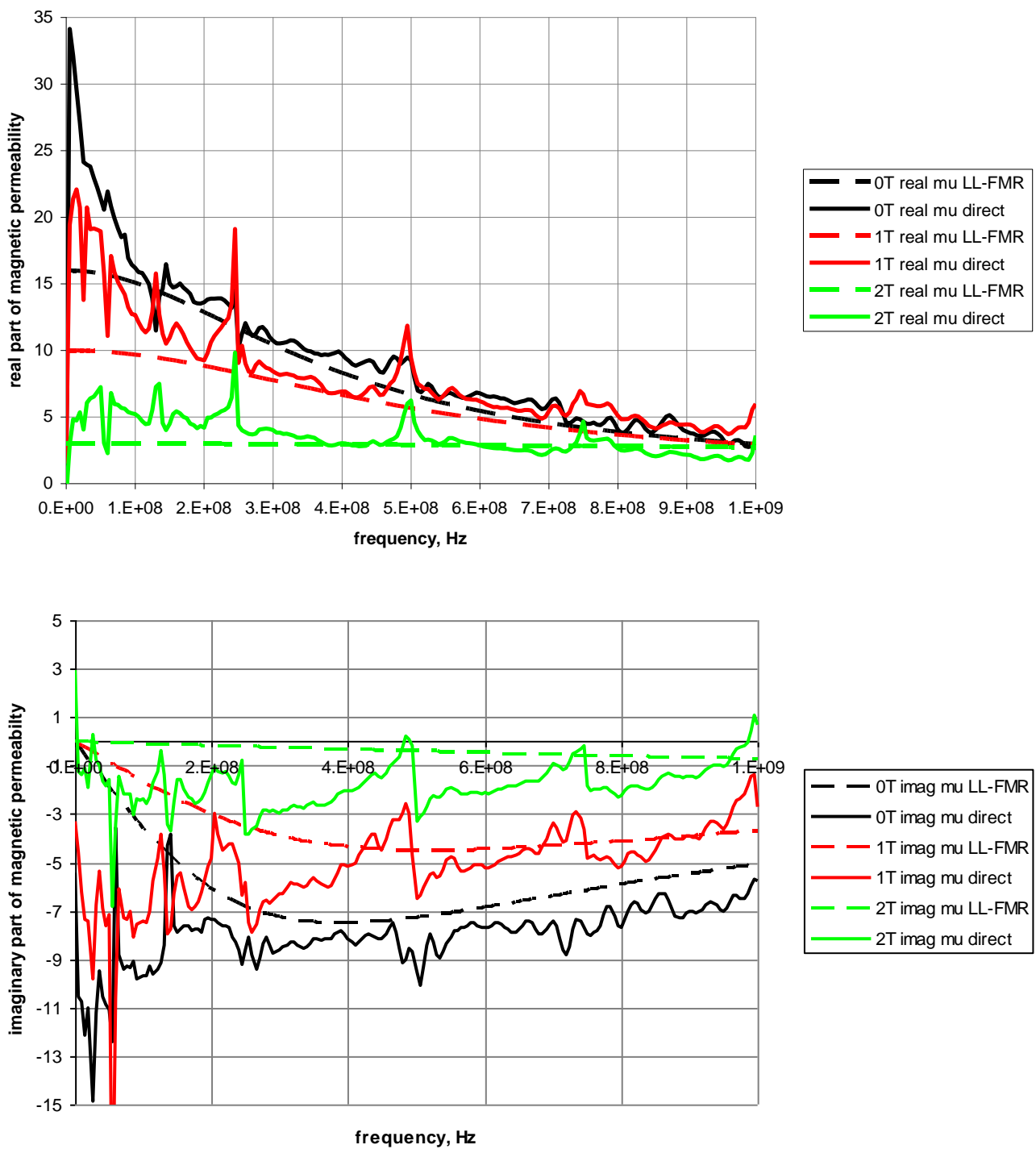

Fig.15 Calculated magnetic permeability of steel in magnetic field, normal to strip plane; solid line – direct calculations, dash line – LL-FMR formula fitting.

The determined magnetic permeability of the steel in normal magnetic fields is presented on the fig. 15. As anticipated both real and imaginary parts decreased with higher magnetic field. In addition, one can see that irregularities are still presented, especially in the region of relatively small frequencies (because in this region the losses are pretty low). However, our analysis showed that these irregularities were much bigger, if we didn't use reference copper data at each value of the magnetic field.

### *4.2.3 Steel measurements in parallel magnetic field*

As mentioned earlier, we had to use epoxy glue in order to tightly attach steel strip to dielectric. This glue affected measurements. That's why in the analysis of this data it is necessary to have a reference steel data in zero magnetic field. Once the resistance per unit length in zero magnetic field is known, the resistances at other magnetic field can be easily calculated:

$$\tau_{ref} = \tau_c \sqrt{1 + \frac{R_{ref}}{j\omega L}}, \quad \tau = \tau_c \sqrt{1 + \frac{R}{j\omega L}} \qquad (17)$$

Time delay can be determined by the numerical procedure, stated in the previous section. Assuming $\tau_c$ to be determined only by geometric and dielectric properties (not by the strip conductor), which is pretty much true, one can express it from the first equation of (17) and substitute it into the second equation of (17). Then $R$ can be easily calculated. For estimation of inductance per unit length $L$ any data can be used, because this parameter is determined only by geometry, which was quite the same in all measurements.

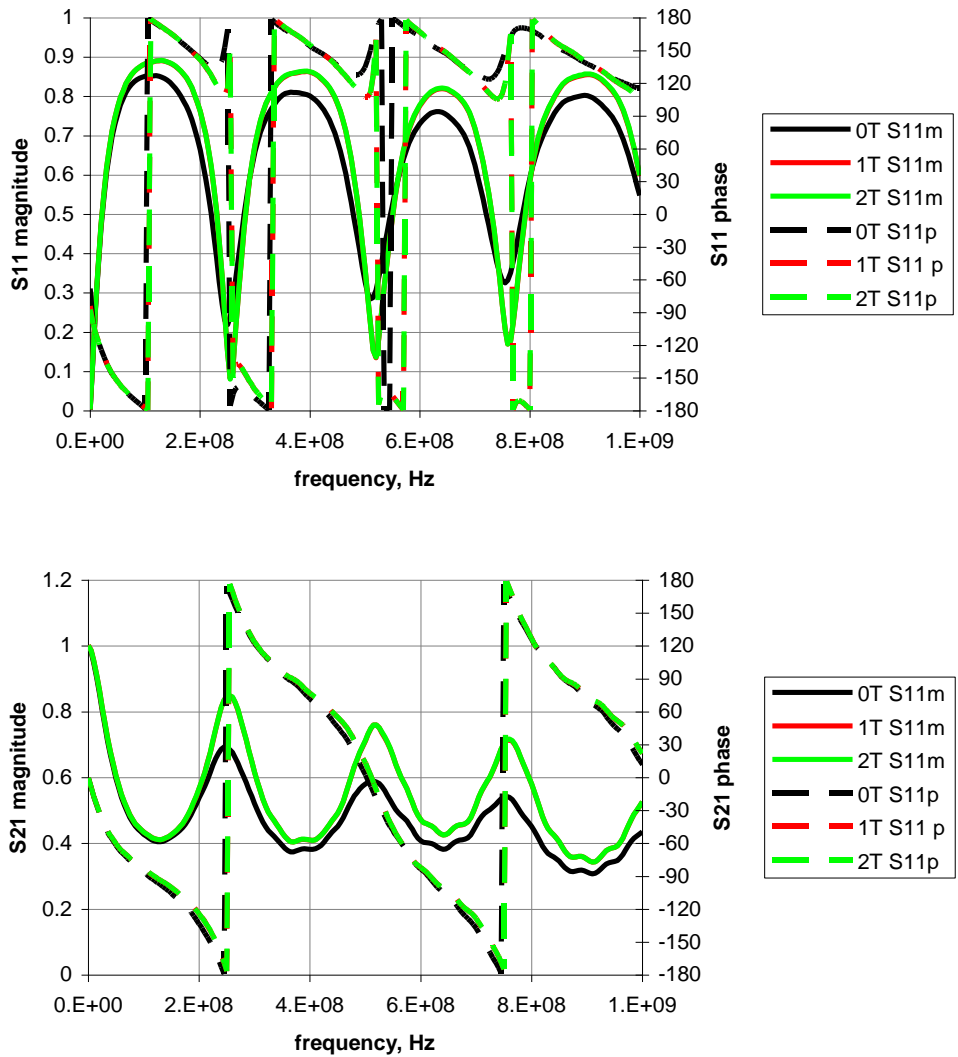

Fig.16. S-parameters of steel microstrip line in different values of magnetic field parallel to the strip plane.

Fig. 16 shows the difference between steel in different values of parallel magnetic field. Such a difference can be seen in the calculated values of magnetic permeability (fig. 17).

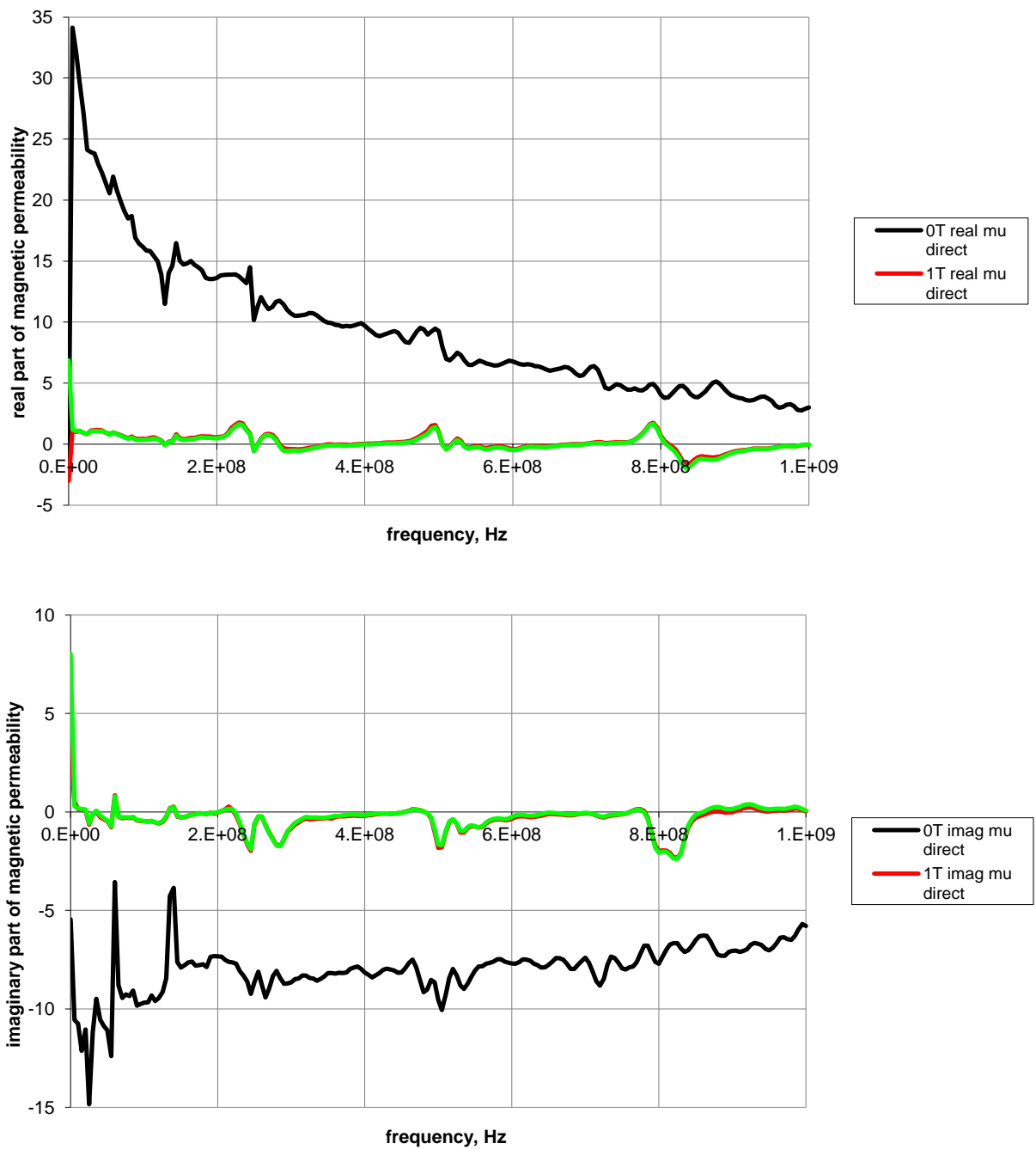

Fig.17 Calculated magnetic permeability of steel in magnetic field, parallel to strip plane.

This data shows a dramatic decrease in magnetic permeability. Several irregularities on this picture are just measuring system errors and don't convey any useful information. No difference between 2T and 1T data means that steel was already in saturation.

The reason for such a big difference between values of magnetic permeability, calculated from data, taken in different orientations of magnetic fields, lies in different boundary conditions. Assuming infinitely wide strip, in the case of magnetic field, normal to the strip plane, the values of B-field are equal inside and outside of the strip in the vicinity of its surface. But in the case of magnetic field, parallel to the strip plane, the same situation appears for H-field. Since by definition $B = \mu\mu_0 H$, the steel reaches saturation in parallel magnetic field much earlier than in normal magnetic field.

## 5. Conclusions

During this project the following goals were achieved:
- The technique for measuring magnetic permeability by analyzing S-parameters is tested and developed.
- Detailed experimental investigation of the problem is carried out, several obstacles, such as errors of the measuring system, are taken into account.
- The estimation of magnetic permeability as a complex function of frequency in the frequency range from several megahertz up to 1 GHz is obtained in different values and orientations of the magnetic field.

## 6. Acknowledgements

I want to acknowledge:
- my advisors Valery Lebedev and Bill Pellico for help in theoretical and experimental work;
- Dallas Heikkinen for providing network analyzer;
- AD and TD for providing all needed equipment;
- PARTI program and Fermilab for providing an outstanding opportunity for making research in such a famous place.